\documentclass[aps,prl,twocolumn,groupedaddress,showpacs]{revtex4-1}
\usepackage{graphicx}% Include figure files
\usepackage{dcolumn}% Align table columns on decimal point
\usepackage{bm}% bold math
\usepackage{amsmath}
\usepackage{epstopdf}

\begin{document}

%\preprint{}

\title{Kennard-Stepanov relation connecting absorption and emission spectra in an atomic gas}



\author{Peter Moroshkin}
\altaffiliation{present address: RIKEN, Center for Emerging Matter
Science, 2-1 Hirosawa, Wako, Saitama 351-0198, Japan}
\email[]{petr.moroshkin@riken.jp}
\author{Lars Weller}
\author{Anne Sa{\ss}}
\author{Jan Klaers}
\author{Martin Weitz}

\affiliation{Institute for Applied Physics, University of Bonn,
Wegelerstr. 8, 53115 Bonn, Germany}

\date{\today}

\begin{abstract}
The Kennard-Stepanov relation describes a thermodynamic,
Boltzmann-type scaling between the absorption and emission
spectral profiles of an absorber, which applies in many liquid
state dye solutions as well as in semiconductor systems. Here we
examine absorption and emission spectra of rubidium atoms in dense
argon buffer gas environment. We demonstrate that the
Kennard-Stepanov relation between absorption and emission spectra
is well fulfilled in the collisionally broadened atomic gas
system. Our experimental findings are supported by a simple
theoretical model.
\end{abstract}


\maketitle


Atomic spectroscopy in the gas phase is an area where electronic
resonances can be investigated extremely well shielded from the
external environment, in which case under linear spectroscopy
conditions the absorption spectral profile of a line matches the
emission spectral profile. This is in striking contrast to many
observations of solid or liquid phase optical spectroscopy, where
the coupling to phonons and other excitations causes a large
variety of optical phenomena \cite{*[{See, \textit{e.g.}:
}][{}]KlingshirnBook}. In some cases, as described by Kennard and
Stepanov \cite{KennardPR1918,StepanovDAN1957}, the coupling to the
environment causes a simple and universal thermodynamic relation
between the absorption and emission lineshapes. This condition can
also be understood to be a consequence of the detailed balance
(Kubo-Martin-Schwinger) condition \cite{BreuerPetruccioneBook}.
The so-called Kennard-Stepanov law states that the spectral
profiles of absorption and emission $\alpha$($\omega$) and
$f$($\omega$) respectively are related by
$f(\omega)$/$\alpha(\omega)\propto\exp(-\frac{\hbar\omega}{k_{B}T})$,
where $T$ denotes the temperature of the sample. This
Boltzmann-type scaling is known to apply for many dye molecules in
liquid solution \cite{LakowiczBook, SawickiPRA1996}, biochemical
photosystems \cite{ZucchelliBBA1995}, as well as for some
semiconductor systems \cite{BandPRA1988}, with observed deviations
being attributed to incomplete thermalization of the excited state
manifold, finite quantum efficiency or internal state conversion.
The Kennard-Stepanov relation is important for spectroscopic
studies and offers prospects for non-contact temperature
determinations. It also plays an important role in recent
experiments on the thermalization of a photon gas by repeated
absorption and emission processes and Bose-Einstein condensation
of photons in a dye microcavity \cite{KlaersN2010}. Indirectly,
the Kennard-Stepanov relation enables the operation of
alkali-vapor lasers in buffer gas broadened systems
\cite{MarkovPRL2002,KrupkePQE2012}.

Here we report the observation of a Kennard-Stephanov scaling
between absorption and emission spectral lineshapes in an atomic
gas. In our experiment, we examine rubidium $D$-lines absorption
and emission spectra at high pressure argon buffer gas conditions.
The typical pressure broadened linewidth reaches a few nanometers,
approaching the thermal energy in frequency units $k_{B}T/\hbar$.
The emission spectral lineshape clearly does not coincide with the
absorption spectrum, but rather exhibits a strong Stokes shift, as
understood from the coupling to the buffer gas reservoir. We
demonstrate that in the dense atomic gas system, despite the
complexity of the individual lineshapes, the ratio between
observed absorption and emission spectral profiles well follows
the frequency dependent thermodynamic scaling predicted by the
Kennard-Stepanov law. Our experimental results are well described
by a theoretical model that is based on the equilibrium between
external atomic degrees of freedom and the internal structure of
alkali-noble gas quasimolecules in the electronic ground and
excited state submanifolds respectively. The agreement between
thermodynamic theory and experiment improves with buffer gas
pressure. In the high pressure regime, the dense atomic buffer gas
system represents a model system for solid and liquid state
systems, well fulfilling the predictions of Kennard and Stepanov.

Before proceeding, we note that it is known that the collisions
between electronically excited state alkali atoms and noble gas
perturbers are surprisingly elastic, so that a large number of
collisions can occur without quenching of the excited state
\cite{SpellerZPA1979}. The interaction of atoms and light near the
line centers is usually well described within the impact limit of
the binary collisional model, which assumes short phase-changing
collisions and the atom-light interaction mainly occuring at times
where the atoms have unperturbed eigenfrequencies. In the impact
limit the predicted lineshapes are symmetric, and the
Kennard-Stepanov relation does not apply. For increased pressures,
as the coupling by the light field experienced during the collision
contributes significantly to the observed spectrum, the impact limit
ceases to be fulfilled. A general description of absorption and
emission probabilities requires a knowledge of the quasimolecular
potential curves of the collision partners, a regime also known as
collisionally aided excitation \cite{AllardRMP1982}. It furthermore
has been noted that frequent atomic collisions can under suitable
conditions lead to a thermalization of quasimolecular states
\cite{HedgesPRA1972,SellJPB2012}.

The Kennard-Stepanov relation can be straightforwardly derived
\cite{McCumberPR1964} for a system with an electronic ground state
$|g\rangle$ and an electronically excited state $|e\rangle$, each of
which are subject to an additional sublevel structure. We assume
that the excited state lifetime is sufficiently long that its
sublevel population (as well as the ground state manifold) acquires
thermal equilibrium. The following derivation follows the treatment
of Sawicki and Knox for rovibrational submanifolds in dye molecules
\cite{SawickiPRA1996}. Our aim is to generalize it to the here
discussed problem of a collisionally broadened atomic system. The
ratio between emission and absorption coefficient at a particular
frequency $\omega$ is
\begin{equation}
\frac{f(\omega)}{ \alpha (\omega)} \propto \frac{\int
g'(E')A(E',\omega)W_{e}(E')\text{d}E'}{\int
g(E)B(E,\omega)W_{g}(E)\text{d}E}, \label{eq1}
\end{equation}
where $E$ and $E'$ are the energies in the electronic ground and
excited state manifolds respectively and $g(E)$ and $g'(E')$ are the
corresponding densities of states. Further, $B(E,\omega)$ and
$A(E',\omega)$ denote the Einstein coefficients for absorption and
spontaneous emission respectively. Note that the integrals in
Eq.~(\ref{eq1}) must be carried out over both the bound and the free
eigenstates, and include quasimolecular states as well as
\textit{e.g.} the atomic fine structure. In thermal equilibrium, the
probabilities of occupation of sublevels are given by the Boltzmann
factors $W_{g}(E)=(1/Q_{g}) \exp(-E/k_{B}T)$ and
$W_{e}(E')=(1/Q_{e}) \exp(-E'/k_{B}T)$ in the lower and upper
electronic states respectively, where $Q_{g}$ and $Q_{e}$ are the
corresponding partition functions. The Einstein $A-B$ relation here
takes the form
\begin{equation}
g'(E')A(E',\omega)\text{d}E' = \frac{2 \hbar \omega^{3}}{\pi
c^{2}}g(E)B(E,\omega)\text{d}E. \label{eq2}
\end{equation}
Substituting this equation and the occupation probabilities
$W_{g}(E)$, $W_{e}(E')$ into eq. (\ref{eq1}) and using
$\hbar\omega_{0}+E'=\hbar\omega+E$, where $\omega_{0}$ denotes the
unperturbed atomic resonance frequency, yields the Kennard-Stepanov
relation
\begin{equation}
\frac{f(\omega)}{ \alpha (\omega)} \propto
\exp\left[-\frac{\hbar(\omega-\omega_{0})}{k_{B}T}\right],
\label{eq3}
\end{equation}
which also can be written in the form
\begin{equation}
ln \left[\frac{\alpha (\omega)}{f(\omega)} \right]=
\frac{\hbar\omega}{k_{B}T} + C, \label{eq4}
\end{equation}
with $C$ being a constant. We note that the same result as above can
also be obtained in a model more closely following standard
collisional theory \cite{AllardRMP1982}, if we assume that the
atomic motion is sufficiently slow so that the quasistatic
approximation applies. The corresponding derivation is given in the
Supplemental Information.

\begin{figure}[tbp]
\includegraphics[width=\columnwidth]{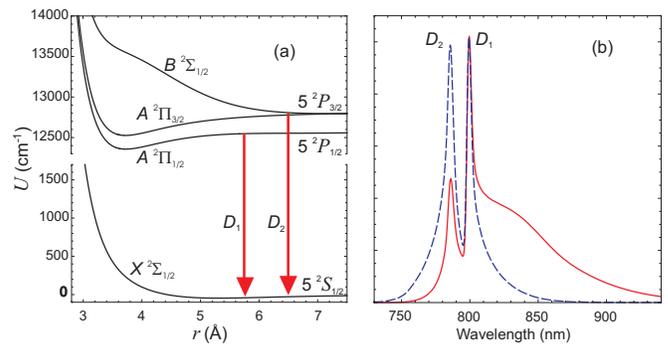}
\caption{(color online) (a) Theoretical rubidium-argon interaction
potentials \cite{DhiflaouiJPCA2012}, as relevant for collisional
broadening of the rubidium $D$-lines. (b) Calculated fluorescence
(solid red line) and absorption (dashed blue line) spectra of
rubidium atoms perturbed by collisions with argon buffer gas atoms.
$T$ = 400 K, $p$ = 180 bar.} \label{fig:Theory}
\end{figure}

Fig. \ref{fig:Theory}(a) gives calculated quasistatic rubidium-argon
molecular potential curves, as derived by ab-initio methods in
Ref.\cite{DhiflaouiJPCA2012}. For large interatomic distance, the
shown curves asymptotically approach the energy of rubidium atoms in
the 5$S_{1/2}$, 5$P_{1/2}$, and 5$P_{3/2}$ electronic states. Using
such potential curves, we have for a more complete characterization
of our system calculated both absorption and emission lineshapes of
the rubidium $D$-lines in the high pressure argon buffer gas regime,
following standard methods of a unified line broadening theory
\cite{AllardRMP1982,RoyerPRA1980,MoroshkinPRA2013}, see the
Supplemental Information. Typical calculated results for the
rubidium $D$-lines absorption and emission spectra are shown in Fig.
\ref{fig:Theory}(b) for a buffer gas pressure of 180 bar.

For an experimental determination of rubidium absorption and
emission in dense buffer gas environment, we use a stainless steel
pressure cell ($\approx$2 cm$^{3}$ inner volume) with sapphire
windows to provide optical access. The cell is filled with
rubidium metal and heated to temperatures in the range of 400 -
600 K, which provides a vapor-pressure limited rubidium density in
the range of 3$\times$10$^{13}$ - 5$\times$10$^{16}$ cm$^{-3}$. We
use argon buffer gas pressures in the range of 20 - 200 bar,
corresponding to densities of order 10$^{21}$ cm$^{-3}$. Rubidium
atoms are excited with a beam from a Ti:sapphire laser, which can
be tuned in a broad range around the rubidium $D_{1}$ and $D_{2}$
lines (with line centers at 795 nm and 780 nm respectively). The
typical used beam power is 0.5 - 1.0 W on a 3 mm beam diameter
($1/e^{2}$ width). We collect the laser-induced fluorescence
emitted in perpendicular direction to the excitation laser beam
and analyze the fluorescence spectrally with a grating
spectrometer of 1 nm resolution. At a typical buffer gas pressure
of 180 bar, the estimated time between two successive collisions
is around 1 ps \cite{OttingerPRA1975}, which is more than four
orders of magnitude below the 27 ns natural lifetime of the
rubidium 5$P$ excited state.

To allow for the investigation of atomic samples with low optical
density, absorption spectra have been obtained by recording the
fluorescence power integrated over the emission spectrum and
plotting it versus the incident laser wavelength, which gives a
signal that is proportional to the absorption coefficient. In this
way the rubidium density (and correspondingly the optical density)
could be kept low enough to suppress reabsorption in the sample.
This technique assumes that the quantum efficiency is near unity (or
at least independent from wavelength), as is well established from
\cite{SpellerZPA1979, VoglN2009}. By determining the fluorescence
yield at different optical beam powers, we have verified in
additional measurements that the laser intensity is low enough to
avoid saturation.

\begin{figure}[tbp]
\includegraphics[width=0.7\columnwidth]{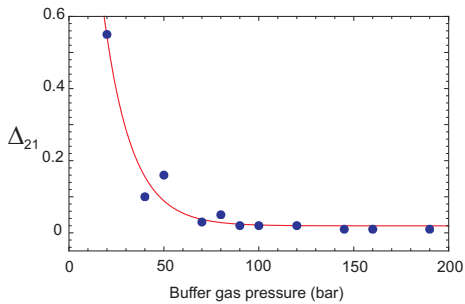}
\caption{(color online) Difference of intensity ratios of
fluorescence signals collected at the rubidium $D_{2}$ and
$D_{1}$-lines
$\Delta_{21}$=$f_{\lambda_{2}}(D_{2})/f_{\lambda_{2}}(D_{1})-
f_{\lambda_{1}}(D_{2})/f_{\lambda_{1}}(D_{1})$, averaged over
excitation wavelength ranges $\lambda_{2}$=760-780 nm and
$\lambda_{1}$=800-820 nm respectively versus the buffer gas
pressure.} \label{fig:D2OverD1}
\end{figure}

In initial measurements, by recording fluorescence spectra at
different incident laser wavelengths, we have verified that the
spectral profile of the emitted fluorescence is, at least in the
high pressure buffer gas regime, largely independent of the
wavelength of the excitation beam. Fig. \ref{fig:D2OverD1} shows the
dependence of the difference of the observed relative fluorescence
signals at the positions of the rubidium $D_{2}$ and $D_{1}$-lines
$f_{\lambda_{2}}(D_{2})/f_{\lambda_{2}}(D_{1})-
f_{\lambda_{1}}(D_{2})/f_{\lambda_{1}}(D_{1})$ averaged over the
excitation wavelengths $\lambda_{2}$=760-780 nm and
$\lambda_{1}$=800-820 nm respectively as a function of the buffer
gas pressure. As in the lower (upper) quoted wavelength range
predominantly the $5P_{3/2}$($5P_{1/2}$)-state is excited, this
difference gives a measure for the redistribution between the 5$P$
fine structure components in collisions with argon atoms. The
difference rapidly drops with buffer gas pressure, and e.g. at 70
bar is at 4$\%$ and further reduces at higher pressures. Similarly,
also the finer lineshape details become independent of the
excitation wavelength, a conclusion in agreement with earlier
experiments of our group in the context of collisional
redistribution laser cooling of rubidium-argon gas mixtures in the
high pressure regime \cite{VoglN2009,VoglJMO2011}, see
\cite{LakowiczBook} for previous work in dye solutions. The
redistribution of fluorescence is attributed to the frequent
collisions of rubidium atoms with argon atoms in the dense gas
system leading to redistribution within the upper electronic state
quasimolecular manifold, with the atoms in the following 'losing
memory' of the value of the exciting laser frequency.

\begin{figure}[tbp]
\includegraphics[width=\columnwidth]{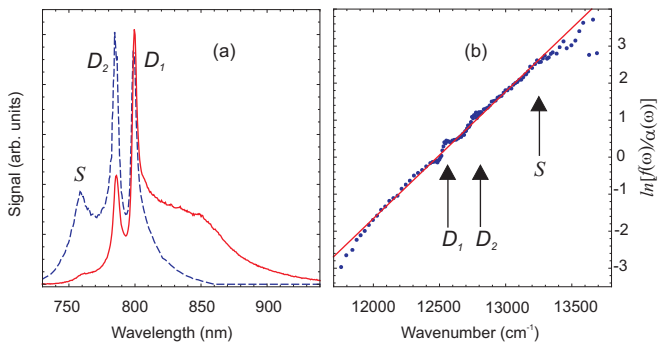}
\caption{(color online) (a) Experimental fluorescence (solid red
line) and absorption (dashed blue line) spectra of the rubidium
$D$-lines, in the presence of argon buffer gas; $p$ = 180 bar, $T$ =
420 K. (b) Corresponding Kennard-Stepanov plot, showing the
logarithm of the ratio of the observed absorption and the emission
spectral profile versus the optical wavenumber. The dots are the
experimental data and the solid line is a guide for the eye with a
slope $hc/k_{B}T$. The vertical arrows indicate the positions of the
$D_{1}$ and $D_{2}$ transitions and of the blue satellite (labeled
$S$) respectively.} \label{fig:Experiment420K}
\end{figure}

Fig.~\ref{fig:Experiment420K}(a) shows typical absorption (dashed
blue line) and fluorescence (solid red line) spectra recorded at a
temperature of 420 K and the buffer gas pressure of 180 bar. The
profile in both cases shows two relatively sharp peaks, the line
cores, near the positions of the rubidium $D_{1}$ and $D_{2}$ lines
respectively on top of a very broad spectral wing, spanning nearly
the complete shown spectral range of 730 - 940 nm. Noticeably, the
line asymmetry of the absorption and emission spectra is quite
opposite, with the absorption (emission) spectra having a more
pronounced short (long) wavelength wing respectively. One may argue
that the spectra thus roughly fulfill the mirror symmetry known
\textit{e.g.} from dye molecules in liquid solution
\cite{LakowiczBook}. Further, the relative peak heights differ in
the two spectra. This is consistent with the predictions for a
coupling of the alkali-noble gas quasimolecular states to the
thermal environment in the dense gas ensemble. Fig.
\ref{fig:Experiment420K}(b) shows the logarithm of the ratio of the
corresponding absorption and the emission spectral profiles versus
the optical wavenumber (which is directly proportional to
frequency). In this representation, the Kennard-Stepanov relation
predicts a linear dependence, see Eq. \ref{eq4}, which when plotted
versus wavenumber has a slope of $hc/k_{B}T$. The dots in Fig.
\ref{fig:Experiment420K}(b) are the corresponding experimental data
of Fig. \ref{fig:Experiment420K}(a), and the solid red line has a
slope of $hc/k_{B}T$, with $T$ corresponding to the cell
temperature. The good linearity of the experimental data, see the
agreement with the red line, suggests that the Kennard-Stepanov law
is well fulfilled in the observed spectral range. This is especially
amazing when taking into account the complex individual lineshapes
of the absorption and emission spectra, in a regime far from the
impact limit. Remaining deviations from the linear dependence are
seen in the vicinity of the $D$-lines resonance frequencies and at
extreme red and blue detunings. The latter is attributed to
technical limitations since in those spectral ranges the absorption
(fluorescence) signal is already very low.

\begin{figure}[tbp]
\includegraphics[width=\columnwidth]{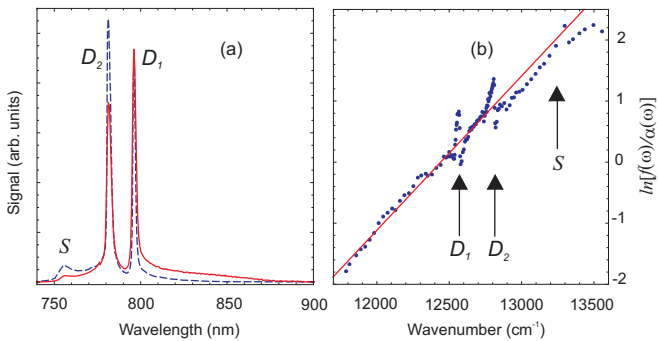}
\caption{(color online) Same as Fig. 3, but for $p$ = 70 bar
buffer gas pressure and $T$ = 570 K. (a) Experimental fluorescence
(solid red line) and absorption (dashed blue line) spectra of
rubidium-argon gas mixture; (b) the corresponding Kennard-Stepanov
plot.} \label{fig:Experiment570K}
\end{figure}

Fig. \ref{fig:Experiment570K}(a) shows absorption and emission
spectra at somewhat lower buffer gas density, with $T$ = 570 K and
$p$ = 70 bar. The cores of both rubidium $D$-lines have a smaller
linewidth, and moreover the spectral wings are less pronounced.
Fig. \ref{fig:Experiment570K}(b) gives the logarithm of the ratio
of the corresponding absorption and the emission profiles, which
again shows a linear scaling that is in a good agreement with a
slope relating to the cell temperature, as predicted by the
Kennard-Stepanov law. We point out that the linewidths of the
$D$-line cores in Fig. \ref{fig:Experiment570K}(a) are only a
factor 2-3 larger than the spectral resolution of our
spectrometer, so that the observed lineshape of the emission
spectrum will thus already have instrumental contributions.
Correspondingly, the visible spectrally sharp substructure near
the $D$-lines resonances in Fig. \ref{fig:Experiment570K}(b) is
still consistent with being purely instrumental. On the other
hand, the deviations from the linear dependence at this lower
buffer gas pressure are also larger far from resonance than in the
case of the 180 bar pressure data. This shows the importance of a
large collisional rate for the thermalization. In other
measurements, we have observed rough agreement with the
Kennard-Stepanov scaling down to argon buffer gas pressures of 40
bar.

The improved agreement of the ratio of absorption and emission with
the Kennard-Stepanov law for larger pressures is attributed to the
role of three-body collisions in the course of thermalization. This
is understood in terms of a redistribution within (quasi-)bound
states of the excited electronic manifold (see the potential minima
of the $A ^{2}\Pi_{3/2}$ or $A ^{2}\Pi_{1/2}$ states in Fig. 1(a))
requiring one additional collisional partner beyond the
argon-rubidium pair. Further, note that the thermalization within
the excited state manifold also requires a redistribution between
the rubidium fine structure components. In earlier work, the role of
three-body collisions for fine structure changing collisions has
been pointed out \cite{SellJPB2012}, which becomes increasingly
important for the much larger buffer gas pressures used in the
present work.

A detailed inspection of the data of Fig.
\ref{fig:Experiment570K}(b), recorded at 70 bar buffer gas
pressure shows that the data at higher wavenumbers beyond the
$D_{2}$ line has the same slope than the data recorded at lower
frequencies, but a different offset. The observed change in offset
could be explained by assuming incomplete redistribution among the
rubidium fine structure levels for the 70 bar data, see also Fig.
\ref{fig:D2OverD1}. In contrast, the 180 bar data of Fig.
\ref{fig:Experiment420K}(b) well follows a single straight line,
both at the positions of the blue and the red wings. The
redistribution results in an enhanced population of the
energetically lower 5$P_{1/2}$ state, and a corresponding decrease
of the $D_{2}$-line emission, see Fig.
\ref{fig:Experiment420K}(a).

We next briefly discuss the agreement of the individual absorption
and emission spectra with our calculated line shape spectra, as were
shown in Fig. \ref{fig:Theory}(b). We point out that the spectral
widths of the line cores and their pressure shifts have been
investigated in some detail in early absorption spectroscopy studies
\cite{ChenRMP1957,WoerdmanPRA1996}. Our experimental data for the
individual absorption and emission spectra, see Figs.
\ref{fig:Experiment420K}(a) and \ref{fig:Experiment570K}(a), exhibit
qualitative agreement with the predictions, as shown in Fig.
\ref{fig:Theory}(b), with the largest deviation probably being the
blue satellite near 750 nm, which is seen in the experimental data
(labeled as ($S$)) but not visible in the theory spectrum. This
satellite is attributed \cite{SUP} to a kink in the $B
^{2}\Sigma_{1/2}$ potential curve near $r$ = 3.97 {\AA}; see Fig.
\ref{fig:Theory}(a). In general, while lineshape calculations are a
nontrivial task in the high pressure buffer gas regime, conclusions
from statistical theory, as the Kennard-Stepanov relation, here
apply with increasing accuracy.

To conclude, we have demonstrated that the Kennard-Stepanov
relation, a thermodynamic frequency dependent scaling between
absorption and emission spectra, is well fulfilled in observed
rubidium $D$-lines spectra at high argon buffer gas pressures. In
general, the agreement improves at larger buffer gas pressures,
which is attributed to the role of three-body processes in both
the thermalization of alkali-noble gas quasimolecular states and
fine-structure changing collisions.

For the future, the system can serve as a model for the study of
system-reservoir interactions. We expect that these findings add to
the development of alkali-lasers \cite{MarkovPRL2002,KrupkePQE2012}
as well as to studies of collisional redistribution laser cooling of
dense gases \cite{VoglN2009,VoglJMO2011}. For non-contact
temperature determinations in dense gases by means of the
Kennard-Stepanov relation, auxiliary optical transitions with low
oscillator strengths, as the blue 5$S$ - 6$P$ transition in the
rubidium atomic system, can be investigated to avoid reabsorption
effects. Furthermore, Kennard-Stepanov spectroscopy in dense gas
systems may allow for atom-based determinations of the Boltzmann
constant. A different perspective includes the thermalization of
photon gases in dense pressure broadened gas samples within an
optical microcavity. This may enable the operation of photon
condensates in different spectral regimes, as the vacuum ultraviolet
when using optical transitions in noble gas atoms, as the
$5\text{p}^{6}(^{1}\text{S}_{0})-5\text{p}^{5}(^{2}\text{P}_{3/2})6\text{s}$
transition near 149 nm from the ground state to the lowest excited
state of the xenon atom for photon thermalization.

\section{Supplemental Information}\label{seq:Supplemental}

We begin by giving a derivation of the Kennard-Stepanov relation
using a model based on the well-known formalism of the quasistatic
line-broadening theory \cite{HedgesPRA1972,AllardRMP1982}. The
presence of a buffer gas atom at a distance $r$ from the absorbing
(radiating) atom causes a shift of the resonance frequency
$\omega-\omega_{0}=(U_{e}(r)-U_{g}(r))/\hbar=\varDelta U(r)$, where
$U_{g}(r)$ and $U_{e}(r)$ are the electronic ground and excited
state potential curves respectively (assumed for simplicity to be
isotropic). Further, let $\rho_{g}(r)$ and $\rho_{e}(r)$ denote the
probabilities to find a buffer gas atom at a distance $r$ relatively
to the absorber (emitter) atom in the ground and excited state
respectively. The absorption coefficient at a frequency $\omega$ is
now proportional to the probability to find a buffer gas atom in the
electronic ground state at a distance $r$ relatively to the
absorber:
\begin{equation}
\alpha(\omega)  \propto \rho_{g}(r) 4 \pi r^{2} g_{g} B(r, \omega)
\left(\frac{\text{d} \varDelta U}{\text{d}r}\right)^{-1}.
\label{eq1}
\end{equation}
With a corresponding expression for the emission line shape, we can
rewrite Eq. (\ref{eq1}) of the main text as
\begin{equation}
\frac{f(\omega)}{\alpha(\omega)}  \propto \frac{\rho_{e}(r)4\pi
r^{2} g_{e} A(r, \omega) \left(\frac{\text{d} \varDelta
U}{\text{d}r}\right)^{-1}}{\rho_{g}(r) 4 \pi r^{2} g_{g} B(r,
\omega) \left(\frac{\text{d} \varDelta U}{\text{d}r}\right)^{-1}}.
\label{eq2}
\end{equation}
The Einstein $A-B$ relation now takes the form
\begin{equation}
g_{e} A(r, \omega) = \frac{2 \hbar \omega^{3}}{\pi c^{2}}g_{g} B(r,
\omega). \label{eq3}
\end{equation}
In thermal equilibrium, the atomic density distribution in the
ground and electronically excited states respectively are
\begin{equation}
\rho_{g,e}(r) = \rho_{0} \exp \left[-\frac{U_{g,e}(r)}{k_{B}T}
\right], \label{eq4}
\end{equation}
with $\rho_{0}$ chosen such that the buffer gas atomic distribution
is normalized. In the above Eq. (\ref{eq2}), we have not reduced the
factor (d$\varDelta U$/d$r)^{-1}$, which arises from the dependence
of the transition frequency on the interatomic distance. In the
limit of the validity of the quasistatic approach, \textit{i.e.} for
d$\varDelta U$/d$r$ differing from zero, the described factor can be
reduced and we find
\begin{equation}
\frac{f(\omega)}{\alpha (\omega)}  \propto \exp
\left[-\frac{(U_e(r)-U_g(r))}{k_{B}T} \right]
=\exp\left[-\frac{\hbar(\omega-\omega_0)}{k_{B}T} \right],
\label{eq5}
\end{equation}
which again verifies the Kennard-Stepanov relation, see eq.3 of the
main text.

For calculations of absolute lineshapes of rubidium atoms at high
argon pertuber densities, as shown in Fig. 1(b), a unified line
broadening theory is used. The applied technique is also known as
the Anderson-Talman-theory \cite{AllardRMP1982}. In brief,
absorption and emission spectra are obtained by a Fourier transform
of an autocorrelation function of the oscillating atomic dipole,
$\Phi(s)$:
\begin{equation}
\Phi(s) = \left\langle \exp\left[-\frac{i}{\hbar} \sum_{k =
1}^{N}\int_{t}^{t + s} \Delta U(\textbf{r}_{k}(t'))
dt'\right]\right\rangle_{t}. \label{eq:CorrelationFunction1}
\end{equation}
Here, it is assumed that the rubidium atom interacts with $N$ argon
atoms (perturbers) occupying positions \textbf{r}$_{k}$ and the
angular brackets denote averaging over initial collision geometries.
It is also assumed that the perturbers do not interact with each
other and their motion is uncorrelated.
Eq. \ref{eq:CorrelationFunction1} is simplified by neglecting the
motion of individual perturbers and introducing the
$\textbf{r}$-dependent perturber densities defined by eq. 4, which
gives
\begin{equation}
\Phi(s) \simeq \exp\left[-\int \left(1 - \exp\left[-\frac{i}{\hbar}
\Delta U(\textbf{r})s \right]\right)\rho_{g(e)}(\textbf{r})
d\textbf{r}\right]. \label{eq:CorrelationFunction2}
\end{equation}

The lineshapes presented in this paper, as shown in Fig. 1(b), were
computed by using the theoretical \textit{ab initio} Rb-Ar
interaction potentials from \cite{DhiflaouiJPCA2012}. Note that
$\Delta U(\textbf{r})$ and $\rho_{e,g}(\textbf{r})$ are spherically
symmetric in the case of the $5^{2}S_{1/2}$ and $5^{2}P_{1/2}$
states of Rb ($X ^{2}\Sigma_{1/2}$ and $A ^{2}\Pi_{1/2}$ potential
curves) and are anisotropic for the $5^{2}P_{3/2}$ state of Rb, for
which $A ^{2}\Pi_{3/2}$ and $B ^{2}\Sigma_{1/2}$ potential curves
describe Rb-Ar interaction along two orthogonal directions.
Electronically-excited Rb atoms may form a bound state
(quasimolecule) with an Ar atom located in a potential well of the
state $A ^{2}\Pi_{3/2}$ or $A ^{2}\Pi_{1/2}$.
A contribution of those bound states is approximated as in
\cite{HedgesPRA1972} by allowing $U_{e}(\textbf{r})$ to take
negative values in the region of the potential well. A more detailed
account of our calculations will be given elsewhere.

The largest disagreement of our experimental lineshapes with
calculations is at around 750 nm wavelength, where the experimental
data, see e.g. Fig. 3(a), show a blue satellite (labeled as $S$)
that in the calculated spectra is, depending on the buffer gas
pressure, either weaker than in the experimental spectra or, as in
the theory spectrum of Fig. 1(b), not seen at all. Such satellites
are well known in the spectra of alkali-rare gas mixtures and other
systems \cite{SzudyPR1996,AllardAA2006}. One expects a satellite at
$\lambda \approx$ 750 nm to arise because of a kink in the $B
^{2}\Sigma_{1/2}$ potential curve at $r$ = 3.97 {\AA} that can be
seen in Fig. 1(a). At this point, the slope of the difference
potential d$\varDelta U$/d$r$ approaches zero and Eq. \ref{eq5}
diverges. The same kind of divergence occurs at large internuclear
separations, where both potential curves are flat. The corresponding
points usually give rise to sharp peaks in the absorption and
emission spectra that are well described by the unified
line-broadening theory.

Our line shape calculations also suggest that at even higher buffer
gas densities than used in the present experiment the probability
that an argon atom remains at very low distance to a rubidium atom
in the electronically excited state manifold approaches unity. This
then leads to a dominating contribution of a quasimolecular band
centered at 850 nm wavelength and a almost complete disappearance of
the sharp $D$-lines spectral peaks.

The experimental absorption spectra have been determined by
measuring series of fluorescence spectra while scanning the incident
laser wavelength, which allows us to use low optical density samples
(see the main text). Each recorded fluorescence spectrum was
normalized to the measured incident laser power. A signal that is
proportional to the absorption was obtained by integrating the
fluorescence yield over the emission wavelength. To suppress a
background from scattered incident laser beam radiation, the
integration range was limited to spectral regions offset from the
excitation laser wavelength. The accuracy of this measurement is
limited by the degree of spectral redistribution. We have verified
that the obtained absorption profile at Ar pressures above 40 bar
does in good accuracy not depend on the choice of the spectral
region in which the fluorescence yield was integrated (complete
redistribution). Fig. 2 summarizes the result of part of these
measurements with regards to the redistribution of the relative
intensities of the rubidium $D_{1}$- and $D_{2}$-lines. In general,
for the absorption measurements the laser wavelength is typically
tuned in 0.5 nm steps. The step size is decreased in the vicinity of
the rubidium $D_{1}$ and $D_{2}$-lines peaks, and increased in the
far wings. Our technique also assumes that the quantum efficiency is
near unity, or at least is independent on the wavelength, as is well
established experimentally for the alkali atom-noble gas mixture
\cite{VoglN2009}.

%\appendix*

\begin{acknowledgments}

We thank H. Berriche for providing numerical data of rubidium-argon
interaction potentials \cite{DhiflaouiJPCA2012}. Financial support
from the DFG (We 1785/15) and the ERC (INPEC) is acknowledged.

\end{acknowledgments}

%\bibliography{RbArSpectra14}

\begin{thebibliography}{27}%
\makeatletter
\providecommand \@ifxundefined [1]{%
 \@ifx{#1\undefined}
}%
\providecommand \@ifnum [1]{%
 \ifnum #1\expandafter \@firstoftwo
 \else \expandafter \@secondoftwo
 \fi
}%
\providecommand \@ifx [1]{%
 \ifx #1\expandafter \@firstoftwo
 \else \expandafter \@secondoftwo
 \fi
}%
\providecommand \natexlab [1]{#1}%
\providecommand \enquote  [1]{``#1''}%
\providecommand \bibnamefont  [1]{#1}%
\providecommand \bibfnamefont [1]{#1}%
\providecommand \citenamefont [1]{#1}%
\providecommand \href@noop [0]{\@secondoftwo}%
\providecommand \href [0]{\begingroup \@sanitize@url \@href}%
\providecommand \@href[1]{\@@startlink{#1}\@@href}%
\providecommand \@@href[1]{\endgroup#1\@@endlink}%
\providecommand \@sanitize@url [0]{\catcode `\\12\catcode
`\$12\catcode
  `\&12\catcode `\#12\catcode `\^12\catcode `\_12\catcode `\%12\relax}%
\providecommand \@@startlink[1]{}%
\providecommand \@@endlink[0]{}%
\providecommand \url  [0]{\begingroup\@sanitize@url \@url }%
\providecommand \@url [1]{\endgroup\@href {#1}{\urlprefix }}%
\providecommand \urlprefix  [0]{URL }%
\providecommand \Eprint [0]{\href }%
\providecommand \doibase [0]{http://dx.doi.org/}%
\providecommand \selectlanguage [0]{\@gobble}%
\providecommand \bibinfo  [0]{\@secondoftwo}%
\providecommand \bibfield  [0]{\@secondoftwo}%
\providecommand \translation [1]{[#1]}%
\providecommand \BibitemOpen [0]{}%
\providecommand \bibitemStop [0]{}%
\providecommand \bibitemNoStop [0]{.\EOS\space}%
\providecommand \EOS [0]{\spacefactor3000\relax}%
\providecommand \BibitemShut  [1]{\csname bibitem#1\endcsname}%
\let\auto@bib@innerbib\@empty
%</preamble>
\bibitem [{\citenamefont {Klingshirn}(2012)}]{KlingshirnBook}%
  \BibitemOpen
  \bibfield  {author} {\bibinfo {author} {\bibfnamefont {C.}~\bibnamefont
  {Klingshirn}},\ }\href@noop {} {\emph {\bibinfo {title} {Semiconductor
  optics}}}\ (\bibinfo  {publisher} {Springer},\ \bibinfo {year}
  {2012})\BibitemShut {NoStop}%
\bibitem [{\citenamefont {Kennard}(1918)}]{KennardPR1918}%
  \BibitemOpen
  \bibfield  {author} {\bibinfo {author} {\bibfnamefont {E.~H.}\ \bibnamefont
  {Kennard}},\ }\href@noop {} {\bibfield  {journal} {\bibinfo  {journal} {Phys.
  Rev.}\ }\textbf {\bibinfo {volume} {11}},\ \bibinfo {pages} {29} (\bibinfo
  {year} {1918})}\BibitemShut {NoStop}%
\bibitem [{\citenamefont {Stepanov}(1957)}]{StepanovDAN1957}%
  \BibitemOpen
  \bibfield  {author} {\bibinfo {author} {\bibfnamefont {B.~I.}\ \bibnamefont
  {Stepanov}},\ }\href@noop {} {\bibfield  {journal} {\bibinfo  {journal}
  {Doklady Akad. Nauk SSSR}\ }\textbf {\bibinfo {volume} {112}},\ \bibinfo
  {pages} {839} (\bibinfo {year} {1957})},\ \bibinfo {note} {{S}oviet Phys. -
  Doklady, 2, 81 (1957)}\BibitemShut {NoStop}%
\bibitem [{\citenamefont {Breuer}\ and\ \citenamefont
  {Petruccione}(2002)}]{BreuerPetruccioneBook}%
  \BibitemOpen
  \bibfield  {author} {\bibinfo {author} {\bibfnamefont {H.~P.}\ \bibnamefont
  {Breuer}}\ and\ \bibinfo {author} {\bibfnamefont {F.}~\bibnamefont
  {Petruccione}},\ }\href@noop {} {\emph {\bibinfo {title} {The theory of open
  quantum systems}}}\ (\bibinfo  {publisher} {Oxford university press,
  Oxford},\ \bibinfo {year} {2002})\BibitemShut {NoStop}%
\bibitem [{\citenamefont {Lakowicz}(1999)}]{LakowiczBook}%
  \BibitemOpen
  \bibfield  {author} {\bibinfo {author} {\bibfnamefont {J.~R.}\ \bibnamefont
  {Lakowicz}},\ }\href@noop {} {\emph {\bibinfo {title} {Principles of
  fluorescence spectroscopy}}}\ (\bibinfo  {publisher} {Kluwer Academic, New
  York},\ \bibinfo {year} {1999})\BibitemShut {NoStop}%
\bibitem [{\citenamefont {Sawicki}\ and\ \citenamefont
  {Knox}(1996)}]{SawickiPRA1996}%
  \BibitemOpen
  \bibfield  {author} {\bibinfo {author} {\bibfnamefont {D.~A.}\ \bibnamefont
  {Sawicki}}\ and\ \bibinfo {author} {\bibfnamefont {R.~S.}\ \bibnamefont
  {Knox}},\ }\href@noop {} {\bibfield  {journal} {\bibinfo  {journal} {Phys.
  Rev. A}\ }\textbf {\bibinfo {volume} {54}},\ \bibinfo {pages} {4837}
  (\bibinfo {year} {1996})}\BibitemShut {NoStop}%
\bibitem [{\citenamefont {Zucchelli}\ \emph {et~al.}(1995)\citenamefont
  {Zucchelli}, \citenamefont {Garlaschi}, \citenamefont {Croce}, \citenamefont
  {Bassi},\ and\ \citenamefont {Jennings}}]{ZucchelliBBA1995}%
  \BibitemOpen
  \bibfield  {author} {\bibinfo {author} {\bibfnamefont {G.}~\bibnamefont
  {Zucchelli}}, \bibinfo {author} {\bibfnamefont {F.~M.}\ \bibnamefont
  {Garlaschi}}, \bibinfo {author} {\bibfnamefont {R.}~\bibnamefont {Croce}},
  \bibinfo {author} {\bibfnamefont {R.}~\bibnamefont {Bassi}}, \ and\ \bibinfo
  {author} {\bibfnamefont {R.~C.}\ \bibnamefont {Jennings}},\ }\href@noop {}
  {\bibfield  {journal} {\bibinfo  {journal} {Biochim. Biophys. Acta}\ }\textbf
  {\bibinfo {volume} {1229}},\ \bibinfo {pages} {59} (\bibinfo {year}
  {1995})}\BibitemShut {NoStop}%
\bibitem [{\citenamefont {Band}\ and\ \citenamefont
  {Heller}(1988)}]{BandPRA1988}%
  \BibitemOpen
  \bibfield  {author} {\bibinfo {author} {\bibfnamefont {Y.~B.}\ \bibnamefont
  {Band}}\ and\ \bibinfo {author} {\bibfnamefont {D.~F.}\ \bibnamefont
  {Heller}},\ }\href@noop {} {\bibfield  {journal} {\bibinfo  {journal} {Phys.
  Rev. A}\ }\textbf {\bibinfo {volume} {38}},\ \bibinfo {pages} {1885}
  (\bibinfo {year} {1988})}\BibitemShut {NoStop}%
\bibitem [{\citenamefont {Klaers}\ \emph {et~al.}(2010)\citenamefont {Klaers},
  \citenamefont {Schmitt}, \citenamefont {Vewinger},\ and\ \citenamefont
  {Weitz}}]{KlaersN2010}%
  \BibitemOpen
  \bibfield  {author} {\bibinfo {author} {\bibfnamefont {J.}~\bibnamefont
  {Klaers}}, \bibinfo {author} {\bibfnamefont {J.}~\bibnamefont {Schmitt}},
  \bibinfo {author} {\bibfnamefont {F.}~\bibnamefont {Vewinger}}, \ and\
  \bibinfo {author} {\bibfnamefont {M.}~\bibnamefont {Weitz}},\ }\href@noop {}
  {\bibfield  {journal} {\bibinfo  {journal} {Nature}\ }\textbf {\bibinfo
  {volume} {468}},\ \bibinfo {pages} {545} (\bibinfo {year}
  {2010})}\BibitemShut {NoStop}%
\bibitem [{\citenamefont {Markov}\ \emph {et~al.}(2002)\citenamefont {Markov},
  \citenamefont {Plekhanov},\ and\ \citenamefont {Shalagin}}]{MarkovPRL2002}%
  \BibitemOpen
  \bibfield  {author} {\bibinfo {author} {\bibfnamefont {R.~V.}\ \bibnamefont
  {Markov}}, \bibinfo {author} {\bibfnamefont {A.~I.}\ \bibnamefont
  {Plekhanov}}, \ and\ \bibinfo {author} {\bibfnamefont {A.~M.}\ \bibnamefont
  {Shalagin}},\ }\href@noop {} {\bibfield  {journal} {\bibinfo  {journal}
  {Phys. Rev. Lett.}\ }\textbf {\bibinfo {volume} {88}},\ \bibinfo {pages}
  {213601} (\bibinfo {year} {2002})}\BibitemShut {NoStop}%
\bibitem [{\citenamefont {Krupke}(2012)}]{KrupkePQE2012}%
  \BibitemOpen
  \bibfield  {author} {\bibinfo {author} {\bibfnamefont {W.~F.}\ \bibnamefont
  {Krupke}},\ }\href@noop {} {\bibfield  {journal} {\bibinfo  {journal} {Prog.
  Quant. Electron.}\ }\textbf {\bibinfo {volume} {36}},\ \bibinfo {pages} {4}
  (\bibinfo {year} {2012})}\BibitemShut {NoStop}%
\bibitem [{\citenamefont {Speller}\ \emph {et~al.}(1979)\citenamefont
  {Speller}, \citenamefont {Staudenmayer},\ and\ \citenamefont
  {Kempter}}]{SpellerZPA1979}%
  \BibitemOpen
  \bibfield  {author} {\bibinfo {author} {\bibfnamefont {E.}~\bibnamefont
  {Speller}}, \bibinfo {author} {\bibfnamefont {B.}~\bibnamefont
  {Staudenmayer}}, \ and\ \bibinfo {author} {\bibfnamefont {V.}~\bibnamefont
  {Kempter}},\ }\href@noop {} {\bibfield  {journal} {\bibinfo  {journal} {Z.
  Phys. A}\ }\textbf {\bibinfo {volume} {291}},\ \bibinfo {pages} {311}
  (\bibinfo {year} {1979})}\BibitemShut {NoStop}%
\bibitem [{\citenamefont {Allard}\ and\ \citenamefont
  {Kielkopf}(1982)}]{AllardRMP1982}%
  \BibitemOpen
  \bibfield  {author} {\bibinfo {author} {\bibfnamefont {N.}~\bibnamefont
  {Allard}}\ and\ \bibinfo {author} {\bibfnamefont {J.}~\bibnamefont
  {Kielkopf}},\ }\href@noop {} {\bibfield  {journal} {\bibinfo  {journal} {Rev.
  Mod. Phys.}\ }\textbf {\bibinfo {volume} {54}},\ \bibinfo {pages} {1103}
  (\bibinfo {year} {1982})}\BibitemShut {NoStop}%
\bibitem [{\citenamefont {Hedges}\ \emph {et~al.}(1972)\citenamefont {Hedges},
  \citenamefont {Drummond},\ and\ \citenamefont {Gallagher}}]{HedgesPRA1972}%
  \BibitemOpen
  \bibfield  {author} {\bibinfo {author} {\bibfnamefont {R.~E.~M.}\
  \bibnamefont {Hedges}}, \bibinfo {author} {\bibfnamefont {D.~L.}\
  \bibnamefont {Drummond}}, \ and\ \bibinfo {author} {\bibfnamefont
  {A.}~\bibnamefont {Gallagher}},\ }\href@noop {} {\bibfield  {journal}
  {\bibinfo  {journal} {Phys. Rev. A}\ }\textbf {\bibinfo {volume} {6}},\
  \bibinfo {pages} {1519} (\bibinfo {year} {1972})}\BibitemShut {NoStop}%
\bibitem [{\citenamefont {Sell}\ \emph {et~al.}(2012)\citenamefont {Sell},
  \citenamefont {Gearba}, \citenamefont {Patterson}, \citenamefont {Byrne},
  \citenamefont {Jemo}, \citenamefont {Lilly}, \citenamefont {Meeter},\ and\
  \citenamefont {Knize}}]{SellJPB2012}%
  \BibitemOpen
  \bibfield  {author} {\bibinfo {author} {\bibfnamefont {J.~F.}\ \bibnamefont
  {Sell}}, \bibinfo {author} {\bibfnamefont {M.~A.}\ \bibnamefont {Gearba}},
  \bibinfo {author} {\bibfnamefont {B.~M.}\ \bibnamefont {Patterson}}, \bibinfo
  {author} {\bibfnamefont {D.}~\bibnamefont {Byrne}}, \bibinfo {author}
  {\bibfnamefont {G.}~\bibnamefont {Jemo}}, \bibinfo {author} {\bibfnamefont
  {T.~C.}\ \bibnamefont {Lilly}}, \bibinfo {author} {\bibfnamefont
  {R.}~\bibnamefont {Meeter}}, \ and\ \bibinfo {author} {\bibfnamefont {R.~J.}\
  \bibnamefont {Knize}},\ }\href@noop {} {\bibfield  {journal} {\bibinfo
  {journal} {J. Phys. B}\ }\textbf {\bibinfo {volume} {45}},\ \bibinfo {pages}
  {055202} (\bibinfo {year} {2012})}\BibitemShut {NoStop}%
\bibitem [{\citenamefont {McCumber}(1964)}]{McCumberPR1964}%
  \BibitemOpen
  \bibfield  {author} {\bibinfo {author} {\bibfnamefont {D.~E.}\ \bibnamefont
  {McCumber}},\ }\href@noop {} {\bibfield  {journal} {\bibinfo  {journal}
  {Phys. Rev.}\ }\textbf {\bibinfo {volume} {136}},\ \bibinfo {pages} {A954}
  (\bibinfo {year} {1964})}\BibitemShut {NoStop}%
\bibitem [{\citenamefont {Dhiflaoui}\ \emph {et~al.}(2012)\citenamefont
  {Dhiflaoui}, \citenamefont {Berriche}, \citenamefont {Herbane}, \citenamefont
  {AlSehimi},\ and\ \citenamefont {Heaven}}]{DhiflaouiJPCA2012}%
  \BibitemOpen
  \bibfield  {author} {\bibinfo {author} {\bibfnamefont {J.}~\bibnamefont
  {Dhiflaoui}}, \bibinfo {author} {\bibfnamefont {H.}~\bibnamefont {Berriche}},
  \bibinfo {author} {\bibfnamefont {M.}~\bibnamefont {Herbane}}, \bibinfo
  {author} {\bibfnamefont {A.~G.}\ \bibnamefont {AlSehimi}}, \ and\ \bibinfo
  {author} {\bibfnamefont {M.~C.}\ \bibnamefont {Heaven}},\ }\href@noop {}
  {\bibfield  {journal} {\bibinfo  {journal} {J. Phys. Chem. A}\ }\textbf
  {\bibinfo {volume} {116}},\ \bibinfo {pages} {10589} (\bibinfo {year}
  {2012})}\BibitemShut {NoStop}%
\bibitem [{\citenamefont {Royer}(1980)}]{RoyerPRA1980}%
  \BibitemOpen
  \bibfield  {author} {\bibinfo {author} {\bibfnamefont {A.}~\bibnamefont
  {Royer}},\ }\href@noop {} {\bibfield  {journal} {\bibinfo  {journal} {Phys.
  Rev. A}\ }\textbf {\bibinfo {volume} {22}},\ \bibinfo {pages} {1625}
  (\bibinfo {year} {1980})}\BibitemShut {NoStop}%
\bibitem [{\citenamefont {Moroshkin}\ \emph {et~al.}(2013)\citenamefont
  {Moroshkin}, \citenamefont {Lebedev},\ and\ \citenamefont
  {Weis}}]{MoroshkinPRA2013}%
  \BibitemOpen
  \bibfield  {author} {\bibinfo {author} {\bibfnamefont {P.}~\bibnamefont
  {Moroshkin}}, \bibinfo {author} {\bibfnamefont {V.}~\bibnamefont {Lebedev}},
  \ and\ \bibinfo {author} {\bibfnamefont {A.}~\bibnamefont {Weis}},\
  }\href@noop {} {\bibfield  {journal} {\bibinfo  {journal} {Phys. Rev. A}\
  }\textbf {\bibinfo {volume} {87}},\ \bibinfo {pages} {022513} (\bibinfo
  {year} {2013})}\BibitemShut {NoStop}%
\bibitem [{\citenamefont {Ottinger}\ \emph {et~al.}(1975)\citenamefont
  {Ottinger}, \citenamefont {Scheps}, \citenamefont {York},\ and\ \citenamefont
  {Gallagher}}]{OttingerPRA1975}%
  \BibitemOpen
  \bibfield  {author} {\bibinfo {author} {\bibfnamefont {C.}~\bibnamefont
  {Ottinger}}, \bibinfo {author} {\bibfnamefont {R.}~\bibnamefont {Scheps}},
  \bibinfo {author} {\bibfnamefont {G.~W.}\ \bibnamefont {York}}, \ and\
  \bibinfo {author} {\bibfnamefont {A.}~\bibnamefont {Gallagher}},\ }\href@noop
  {} {\bibfield  {journal} {\bibinfo  {journal} {Phys. Rev. A}\ }\textbf
  {\bibinfo {volume} {11}},\ \bibinfo {pages} {1815} (\bibinfo {year}
  {1975})}\BibitemShut {NoStop}%
\bibitem [{\citenamefont {Vogl}\ and\ \citenamefont {Weitz}(2009)}]{VoglN2009}%
  \BibitemOpen
  \bibfield  {author} {\bibinfo {author} {\bibfnamefont {U.}~\bibnamefont
  {Vogl}}\ and\ \bibinfo {author} {\bibfnamefont {M.}~\bibnamefont {Weitz}},\
  }\href@noop {} {\bibfield  {journal} {\bibinfo  {journal} {Nature}\ }\textbf
  {\bibinfo {volume} {461}},\ \bibinfo {pages} {70} (\bibinfo {year}
  {2009})}\BibitemShut {NoStop}%
\bibitem [{\citenamefont {Vogl}\ \emph {et~al.}(2011)\citenamefont {Vogl},
  \citenamefont {Sa{\ss}}, \citenamefont {Ha{\ss}elmann},\ and\ \citenamefont
  {Weitz}}]{VoglJMO2011}%
  \BibitemOpen
  \bibfield  {author} {\bibinfo {author} {\bibfnamefont {U.}~\bibnamefont
  {Vogl}}, \bibinfo {author} {\bibfnamefont {A.}~\bibnamefont {Sa{\ss}}},
  \bibinfo {author} {\bibfnamefont {S.}~\bibnamefont {Ha{\ss}elmann}}, \ and\
  \bibinfo {author} {\bibfnamefont {M.}~\bibnamefont {Weitz}},\ }\href@noop {}
  {\bibfield  {journal} {\bibinfo  {journal} {J. Mod. Optics}\ }\textbf
  {\bibinfo {volume} {58}},\ \bibinfo {pages} {1300} (\bibinfo {year}
  {2011})}\BibitemShut {NoStop}%
\bibitem [{\citenamefont {Ch'en}\ and\ \citenamefont
  {Takeo}(1957)}]{ChenRMP1957}%
  \BibitemOpen
  \bibfield  {author} {\bibinfo {author} {\bibfnamefont {S.~Y.}\ \bibnamefont
  {Ch'en}}\ and\ \bibinfo {author} {\bibfnamefont {M.}~\bibnamefont {Takeo}},\
  }\href@noop {} {\bibfield  {journal} {\bibinfo  {journal} {Rev. Mod. Phys.}\
  }\textbf {\bibinfo {volume} {29}},\ \bibinfo {pages} {20} (\bibinfo {year}
  {1957})}\BibitemShut {NoStop}%
\bibitem [{\citenamefont {Woerdman}\ \emph {et~al.}(1996)\citenamefont
  {Woerdman}, \citenamefont {Blok}, \citenamefont {Kristensen},\ and\
  \citenamefont {Schrama}}]{WoerdmanPRA1996}%
  \BibitemOpen
  \bibfield  {author} {\bibinfo {author} {\bibfnamefont {J.~P.}\ \bibnamefont
  {Woerdman}}, \bibinfo {author} {\bibfnamefont {F.~J.}\ \bibnamefont {Blok}},
  \bibinfo {author} {\bibfnamefont {M.}~\bibnamefont {Kristensen}}, \ and\
  \bibinfo {author} {\bibfnamefont {C.~A.}\ \bibnamefont {Schrama}},\
  }\href@noop {} {\bibfield  {journal} {\bibinfo  {journal} {Phys. Rev. A}\
  }\textbf {\bibinfo {volume} {35}},\ \bibinfo {pages} {1183} (\bibinfo {year}
  {1996})}\BibitemShut {NoStop}%
\bibitem [{SUP()}]{SUP}%
  \BibitemOpen
  \href@noop {} {}\bibinfo {note} {See Supplemental Material at ... for
  details.}\BibitemShut {Stop}%
\bibitem [{\citenamefont {Szudy}\ and\ \citenamefont
  {Baylis}(1996)}]{SzudyPR1996}%
  \BibitemOpen
  \bibfield  {author} {\bibinfo {author} {\bibfnamefont {J.}~\bibnamefont
  {Szudy}}\ and\ \bibinfo {author} {\bibfnamefont {W.~E.}\ \bibnamefont
  {Baylis}},\ }\href@noop {} {\bibfield  {journal} {\bibinfo  {journal} {Phys.
  Rep.}\ }\textbf {\bibinfo {volume} {266}},\ \bibinfo {pages} {127} (\bibinfo
  {year} {1996})}\BibitemShut {NoStop}%
\bibitem [{\citenamefont {Allard}\ and\ \citenamefont
  {Spiegelman}(2006)}]{AllardAA2006}%
  \BibitemOpen
  \bibfield  {author} {\bibinfo {author} {\bibfnamefont {N.~F.}\ \bibnamefont
  {Allard}}\ and\ \bibinfo {author} {\bibfnamefont {F.}~\bibnamefont
  {Spiegelman}},\ }\href@noop {} {\bibfield  {journal} {\bibinfo  {journal}
  {Astr. Astrophys.}\ }\textbf {\bibinfo {volume} {452}},\ \bibinfo {pages}
  {351} (\bibinfo {year} {2006})}\BibitemShut {NoStop}%
\end{thebibliography}

%

\end{document}